\pdfoutput=1
\documentclass[fleqn,usenatbib]{mnras}

\usepackage{newtxtext,newtxmath}

\usepackage[T1]{fontenc}

\DeclareRobustCommand{\VAN}[3]{#2}
\let\VANthebibliography\thebibliography
\def\thebibliography{\DeclareRobustCommand{\VAN}[3]{##3}\VANthebibliography}

\usepackage{graphicx}	
\usepackage{amsmath}	

\usepackage[dvipsnames]{xcolor}

\definecolor{pink}{RGB}{255, 20, 147}

\definecolor{carminered}{rgb}{1.0, 0.0, 0.22}

\definecolor{byzantine}{rgb}{1.0, 0.31, 0.0}

\definecolor{blue-violet}{rgb}{0.30, 0.1, 0.89}

\definecolor{amethyst}{rgb}{0.6, 0.4, 0.8}

\definecolor{verde}{rgb}{0.01, 0.53, 0.31}

\definecolor{red}{rgb}{0.0, 0.0, 0.0}

\title[Hubble constant from BBH mergers in AGN]{Determining the Hubble Constant with AGN-assisted Black Hole Mergers}

\author[L. M. B. Alves et al.]{
Lucas M. B. Alves,$^{1}$\thanks{E-mail: lucas.alves@columbia.edu}
Andrew G. Sullivan,$^{2}$
Yang Yang,$^{3}$
Gayathri V.,$^{3,4}$
Zsuzsa Márka,$^{5}$
\newauthor
Szabolcs Márka,$^{1}$
and Imre Bartos$^{3}$\thanks{E-mail: imrebartos@ufl.edu}
\\
$^{1}$Department of Physics, Columbia University in the City of New York, New York, NY 10027, USA\\
$^{2}$Kavli Institute for Particle Astrophysics and Cosmology, Department of Physics, Stanford University, Stanford, CA 94305, USA\\
$^{3}$Department of Physics, University of Florida, PO Box 118440, Gainesville, FL 32611, USA\\
$^{4}$Leonard E. Parker Center for Gravitation, Cosmology, and Astrophysics, University of Wisconsin–Milwaukee, Milwaukee, WI 53201, USA\\
$^{5}$Columbia Astrophysics Laboratory, Columbia University in the City of New York, New York, NY 10027, USA
}

\date{Accepted XXX. Received YYY; in original form ZZZ}

\pubyear{2023}

\begin{document}
\label{firstpage}
\pagerange{\pageref{firstpage}--\pageref{lastpage}}
\maketitle

\begin{abstract}
Gravitational waves from neutron star mergers have long been considered a promising way to measure the Hubble constant, $H_0$, which describes the local expansion rate of the universe. While black hole mergers are more abundantly observed, their expected lack of electromagnetic emission and poor gravitational-wave localization make them less well suited for measuring $H_0$. Black hole mergers within the disks of Active Galactic Nuclei (AGN) could be an exception. Accretion from the AGN disk may produce an electromagnetic signal, pointing observers to the host galaxy. Alternatively, the low number density of AGNs could help identify the host galaxy of $1-5\%$ of mergers. Here we show that black hole mergers in AGN disks may be \textcolor{red}{a sensitive} way to determine $H_0$ with gravitational waves. If \textcolor{red}{$1\%$ ($10\%$) of LIGO's observations occur in AGN disks with identified host galaxies, we could measure $H_0$ with $12\%$ ($4\%$) uncertainty in five years, possibly comparable to} the sensitivity of neutron star mergers \textcolor{red}{and set to considerably improve current gravitational wave measurements}.
\end{abstract}




\section{Introduction}
Multi-messenger gravitational-wave observations represent a valuable, independent probe of the expansion of the universe \citep{1986Natur.323..310S}. The Hubble constant, which describes the rate of expansion, can be measured using type Ia supernovae, giving a local expansion rate of $H_0 = 74.03 \pm 1.42$\,km\,s$^{-1}$\,Mpc$^{-1}$  \citep{2019ApJ...876...85R}. It can also be measured through cosmic microwave background observations focusing on the early universe, which gives a conflicting $H_0 = 67.4\pm 0.5$\,km\,s$^{-1}$\,Mpc$^{-1}$ \citep{Planck:2018vyg}. These results differ at the $4.4\sigma$ level, which may be the signature of new physics beyond our current understanding of cosmology.

Gravitational waves from a compact binary merger carry information about the luminosity distance of the source. To determine $H_0$ one also needs to measure redshift, which is not directly accessible from the source but can be measured from the \textcolor{red}{electromagnetic} spectrum of the merger's host galaxy.  

The identification of the host galaxy typically requires the detection of electromagnetic emission coincident with the limited localization available through gravitational waves. Neutrons star mergers are natural targets for such measurements due to the broad range of electromagnetic emission they produce \citep{Bartos_2013,2012ApJ...746...48M}. The first neutron star merger discovered by LIGO \citep{2015CQGra..32g4001L} and Virgo \citep{2015CQGra..32b4001A}, GW170817, was also observed across the electromagnetic spectrum, and was used to constrain $H_0$ \citep{GW170817_H0,2017ApJ...851L..36G,2019NatAs...3..940H}.

\begin{figure}
\centering
\includegraphics[width=0.46\textwidth]{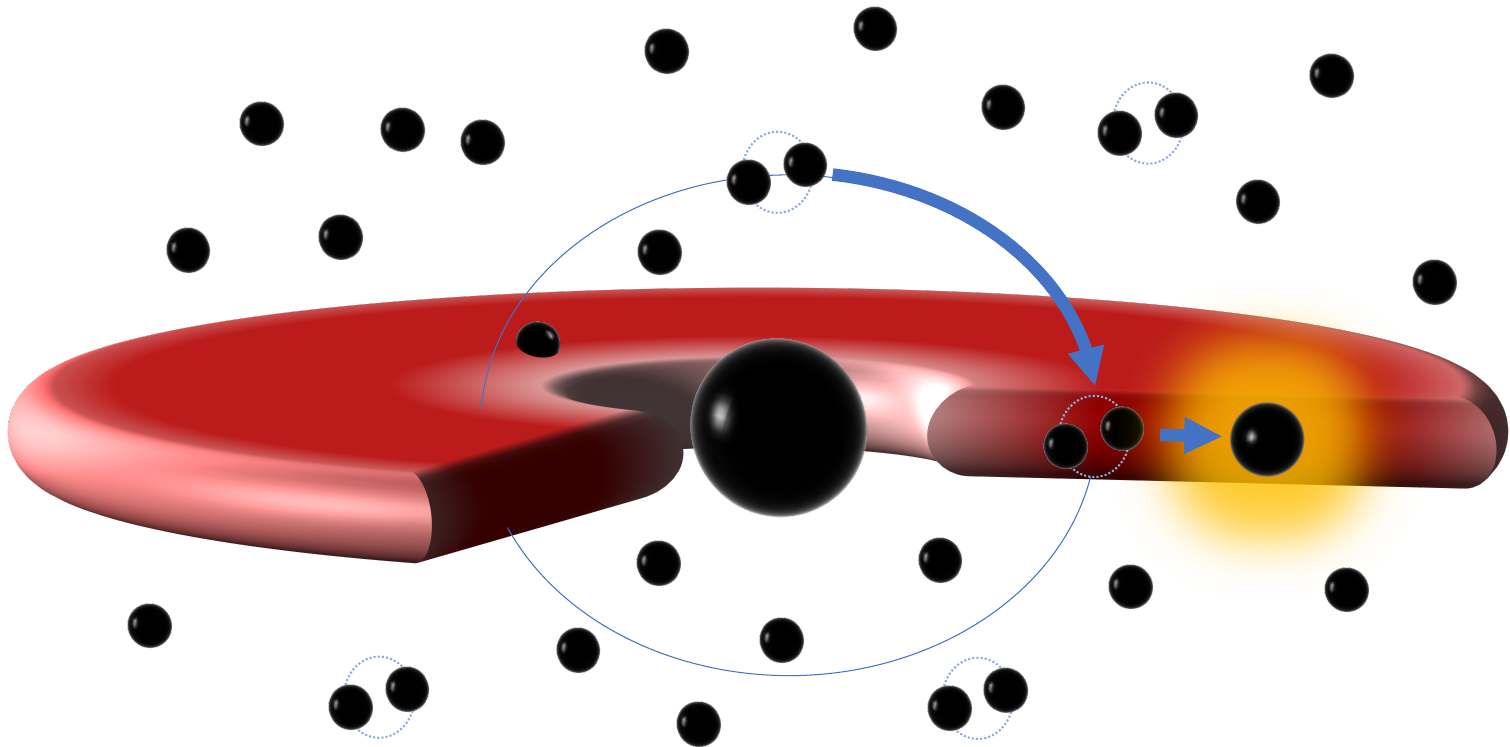}
\caption{{\bf Illustration of black hole mergers in AGN disks}. A binary black hole system migrates into the accretion disk around the central supermassive black hole. It then rapidly merges due to dynamical friction within the disk. Following the merger the remnant black hole produces an optical signal due to accretion from the disk.}
\label{fig:illustration}
\end{figure}

Not all neutron star mergers detected by LIGO-Virgo-KAGRA will have identified electromagnetic counterparts \citep{Hosseinzadeh_2019,2020NatAs.tmp..179A}. Especially more distant events will be difficult to observe electromagnetically due to their weaker flux and poorer gravitational-wave localization. In addition, as neutron star mergers are relatively nearby, the redshift of their host galaxies will be affected significantly by proper motion, introducing a systematic uncertainty in $H_0$ measurements. 

Black hole mergers are detected by LIGO/Virgo at a rate more than an order of magnitude higher than neutrons star mergers \citep{2018arXiv181112907T,KAGRA:2013rdx}. In addition, they are typically detected at much greater distances; therefore, they are less affected by the proper motion of galaxies. Nonetheless, black hole mergers are not generically expected to produce electromagnetic counterparts, limiting their utility for $H_0$ measurements (although see \cite{Del_Pozzo_2012,2019ApJ...876L...7S,2019ApJ...883L..42F,2020arXiv200702943M,2021ApJ...908..215Y}).

Active galactic nuclei (AGN) represent unique environments that facilitate the mergers of black holes and potentially also electromagnetic emission from them \citep{2017ApJ...835..165B,2017MNRAS.464..946S,McKernan_2019}. Galactic centers harbor a dense population of thousands of stellar mass black holes that migrated there through mass segregation \citep{2009MNRAS.395.2127O,2018Natur.556...70H}. These black holes interact with the accretion disk of the central supermassive black hole. Angular momentum exchange drives their orbits to align with the disk plane. The disk can hence become an ultra-dense 2D collection of black holes. The black holes then migrate within the disk. Once two of them get close and form a binary, they rapidly merge due to dynamical friction within the gas or binary single encounters with other nearby objects. Alternatively, a binary black hole (BBH) system formed outside the accretion disk may migrate into it and have its merger sped up due to dynamical friction, as illustrated in Fig. \ref{fig:illustration}. The gas-rich environment of these binaries enables the black holes to accrete and produce electromagnetic radiation.

Recently, such a candidate electromagnetic counterpart has been observed by the Zwicky Transient Facility (ZTF), in coincidence with the black hole merger GW190521 \citep{Graham_2020,GW190521discovery}. In addition, the mass and spin of GW190521 suggest a possible AGN origin \citep{GW190521_properties,Yang_2019_harden,Yang_2020,yang2020black}. While this is the clearest case, some of the other black hole mergers recorded by LIGO/Virgo may have also originated in AGN disks \citep{2023ApJ...942...99G}. 

Compact objects in certain mass ranges are more likely to have an AGN origin. Pair-instability supernovae are thought to prevent black holes from reaching masses in the range from $M_{\mathrm{low}}\approx50M_{\odot}$ to $M_{\mathrm{high}}\approx130M_{\odot}$ via stellar evolution \citep{RevModPhys.74.1015,2007Natur.450..390W}. Recent studies have proposed hierarchical mergers as the formation channel for black holes in the pair-instability mass gap \citep{2020ApJ...890L..20G,10.1093/mnrasl/slaa123,2021ApJ...920L..42G}. These black holes may be concentrated in active environments like AGN, where mergers happen frequently \citep{2023arXiv230214071A,2021PhRvD.104h2003O,2020ApJ...901L..34Y,2021ApJ...908..194T}. The same applies to compact objects in the mass gap between neutron stars and black holes. Objects in this gap have masses of $\sim2.3-3M_{\odot}$, so they cannot be determined to be neutron stars or black holes. This is due to the uncertainty in the maximum neutron star mass arising from the insufficient constraining of the neutron star equation of state \citep{2012ARNPS..62..485L,2016ARA&A..54..401O}. Compact objects in this mass gap are likewise probable to exist in hierarchical mergers and, thus, have AGN origin. More than ten compact objects in either mass gap have been detected so far \citep{LIGOScientific:2021djp,2018arXiv181112907T,Yang_2019,2019arXiv191009528Z,Gayathri_2020,2023ApJ...942...99G,Abbott_2020_GW190814,yang2020black,2020AAS...23533403F,2023arXiv230207284Z}. Events involving such objects are therefore particularly promising for AGN-assisted merger searches.

Even for those mergers where no electromagnetic counterpart is observed, the rarity of AGNs can help the identification of their host galaxies. Counting even the less active Seyfert galaxies, the AGN number density in the local universe is $n_{\rm Seyfert}\approx 0.02$\,Mpc$^{-3}$ \citep{2005AJ....129.1795H}. Considering the contribution of only the more active galactic nuclei, their density is only $n_{\rm AGN}\approx 2\times 10^{-5}$\,Mpc$^{-3}$ \citep{Greene_2007,Greene_2009}. As fig. 6 in \cite{KAGRA:2013rdx} shows, $1-5\%$ of black hole mergers detected by LIGO-Virgo-KAGRA in O4 could be sufficiently well localized such that only a single AGN resides in their 3D localization volume. O5 and next-generation gravitational-wave-detector runs promise more events with such precise 3D volume localization. Even when multiple AGN lie within the 3D volume localization, one may use the structure of the gravitational wave localization to identify more viable AGN candidates.

\section{Bayesian simulation of mergers}

\textcolor{red}{We performed $10^4$ BBH merger injections using BAYESTAR \citep{Singer_2016}. BAYESTAR employs Bayesian inference and detailed models of gravitational wave signals to estimate the distance to binary merger sources based on the observed data and prior information. This approach provides a probabilistic estimate with associated uncertainties, allowing for a better understanding of the properties of the observed events.} 

\textcolor{red}{Our injections are drawn via Monte Carlo methods assuming a broken power-law+peak primary mass distribution and a power-law distributed mass ratio as inferred from the GWTC-3 population analyses \citep{2023PhRvX..13a1048A}. These distance and sky positions are drawn assuming uniform density in comoving volume, similar to our expectation for AGN-assisted mergers \citep{Yang_2020}. Our injected distribution assumes a Hubble constant value $H_0=68$ km s$^{-1}$ Mpc$^{-1}$.} We consider only the case in which the black hole spins are aligned with the binary orbit. 

We simulate the detection of these BBH mergers with a GW detector network comprised of LIGO Hanford and LIGO Livingston. \textcolor{red}{Half of the injections are performed with approximately O4 sensitivity while the other half, with approximately O5 sensitivity \citep{emfollow}. We utilized the SEOBNRv4twoPointFivePN waveform model, a lower frequency of $15$\, Hz, and required that both detectors should be triggered with minimum SNR 3 and net SNR 10.} \textcolor{red}{With BAYESTAR, we obtain the reconstructed luminosity distance $\hat{d}_L$ and its uncertainty $\sigma_{d_{\rm L}}$.}



We assume that a BBH is localized to a single galaxy with known redshift. Since black holes are typically found at Gpc distances, we neglect the peculiar velocity of the host galaxy. Then, in accordance with \cite{2018Natur.562..545C}, the calculated relative uncertainty in $H_0$ is, for a single source, the same as the relative distance uncertainty and, for multiple sources, the root mean square of the relative distance uncertainties. \textcolor{red}{We compute the expected relative uncertainty in $H_0$ from a single BBH for both our O4 and O5 injection populations.}

If the merger occurred within the AGN disk, it could produce a detectable electromagnetic counterpart associated with this event. Such a counterpart can help us identify the host galaxy of the merger and thus the redshift of the event can be determined. The true distance of the event is linked to its redshift by:
\begin{equation}
   d_{\rm L}=\frac{c(1+z)}{H_0}\int_0^{z}\frac{dz'}{E(z')}
   \label{eq:dist}
\end{equation}
with
\begin{equation}
E(z)=\sqrt{\Omega_{\rm r}(1+z)^4+\Omega_{\rm m}(1+z)^3+\Omega_{\rm k}(1+z)^2+\Omega_{\rm \Lambda}}\,.
\end{equation}
Here, $\Omega_{\rm r}$ is the current radiation energy density, $\Omega_{\rm m}$ is the current matter density, $\Omega_{\rm \Lambda}$ is the current dark energy density, and $\Omega_{\rm k}$ is related to the curvature of our universe. We adopted a set of cosmology parameters of $\{H_0, \Omega_{\rm r}, \Omega_{\rm m}, \Omega_{\rm k}, \Omega_{\rm \Lambda}\}=\{68{\rm km\ s^{-1}\ Mpc^{-1}},\ 0,\ 0.306,\ 0,\ 0.694\}$.

The measured luminosity distance $d_{\rm L}$ can also be linked to the redshift $z$ similarly to Eq. \ref{eq:dist} above  but by replacing $H_0$ with the measured Hubble constant $\hat{H}_{\rm 0}$. 
Therefore, we find $\hat{H}_{\rm 0}/H_0=d_{\rm L}/\hat{d}_{\rm L}$. 
%
For this, we assumed that the other cosmological parameters are fixed at the values obtained by Planck 2018 \citep{Planck:2018vyg}.

If we detect multiple AGN-assisted black hole mergers and identify their host galaxies either through their electromagnetic counterpart or through accurate gravitational-wave localization, we will obtain a set of measurements of the Hubble constant, $\{\hat{H}_{\rm 0,1},\hat{H}_{\rm 0,2},...,\hat{H}_{\rm 0,N}\}$. We adopted the arithmetic mean of our simulated values as our overall estimate $H_0$ of the Hubble constant. 
The distribution of $\hat{H_0}$ is asymptotically normal and its standard deviation is $\sigma(\hat{H_0})\propto N^{-1/2}$.

\section{Detection rate of mergers}

With the improving sensitivity of LIGO/Virgo, the inclusion of KAGRA \citep{2021PTEP.2021eA103A}, and later the construction of LIGO-India \citep{LIGOIndia}, the rate of gravitational-wave discoveries will rapidly increase over the next few years \citep{KAGRA:2013rdx}. While currently uncertain, the fraction of detected mergers that originate from AGNs could be $10-50\%$ \citep{Yang_2019_harden}. 

It is also uncertain what fraction of AGN-assisted mergers will have a detected electromagnetic counterpart. While there are theoretical predictions \citep{2017ApJ...835..165B,2017MNRAS.464..946S,McKernan_2019}, the emission process is not yet well understood. To estimate this fraction, we considered the fact that one black hole merger, GW190521 \citep{GW190521discovery}, has a candidate electromagnetic counterpart so far, detected by ZTF \citep{Graham_2020}. In recent analyses, although the significance of the coincidence between GW190521 and this candidate counterpart does not rise to the $3\sigma$ confidence level, there remains a favorable coincidence between the two events \citep{Palmese_2021, 2021CQGra..38w5004A}. For our study, we assumed that this candidate is a real counterpart of an AGN-assisted merger and, furthermore, that it is the only real joint detection, despite recently identified candidates \citep{2023ApJ...942...99G}. Then, the expected fraction of LIGO/Virgo's detections that have electromagnetic counterparts is about 1\%. We conservatively estimated this number by considering all BBH merger events currently in the Gravitational-Wave Transient Catalog as detections \citep{2018arXiv181112907T,LIGOScientific:2021usb,LIGOScientific:2021djp}.

In addition, we may be able to identify the host galaxy of some of the AGN-assisted black hole mergers due to their accurate gravitational-wave localization. As discussed above, $1-5\%$ of all detected binary black hole mergers could be so well localized that a single AGN resides in their localization volume, and $10-50\%$ of all detected mergers should originate from AGNs. This means that $0.1-2.5\%$ of detections would be AGN-assisted and have their host galaxy identified through the gravitational wave signal \textcolor{red}{alone. To estimate the total fraction of AGN-assisted LIGO/Virgo black holes with redshift information, the additional $\sim1\%$ of events with electromagnetic counterparts should be taken into account, although there could be overlap. Furthermore, it is reasonable to assume that the majority of mergers that have a single AGN in their localization volume are AGN-assisted. Considering these, we assumed that $\sim 2\%$ of all black hole mergers detected by Earth-based interferometers are AGN-assisted and have their redshift determined.}

\begin{figure}
\centering  
\includegraphics[width=0.5\textwidth]{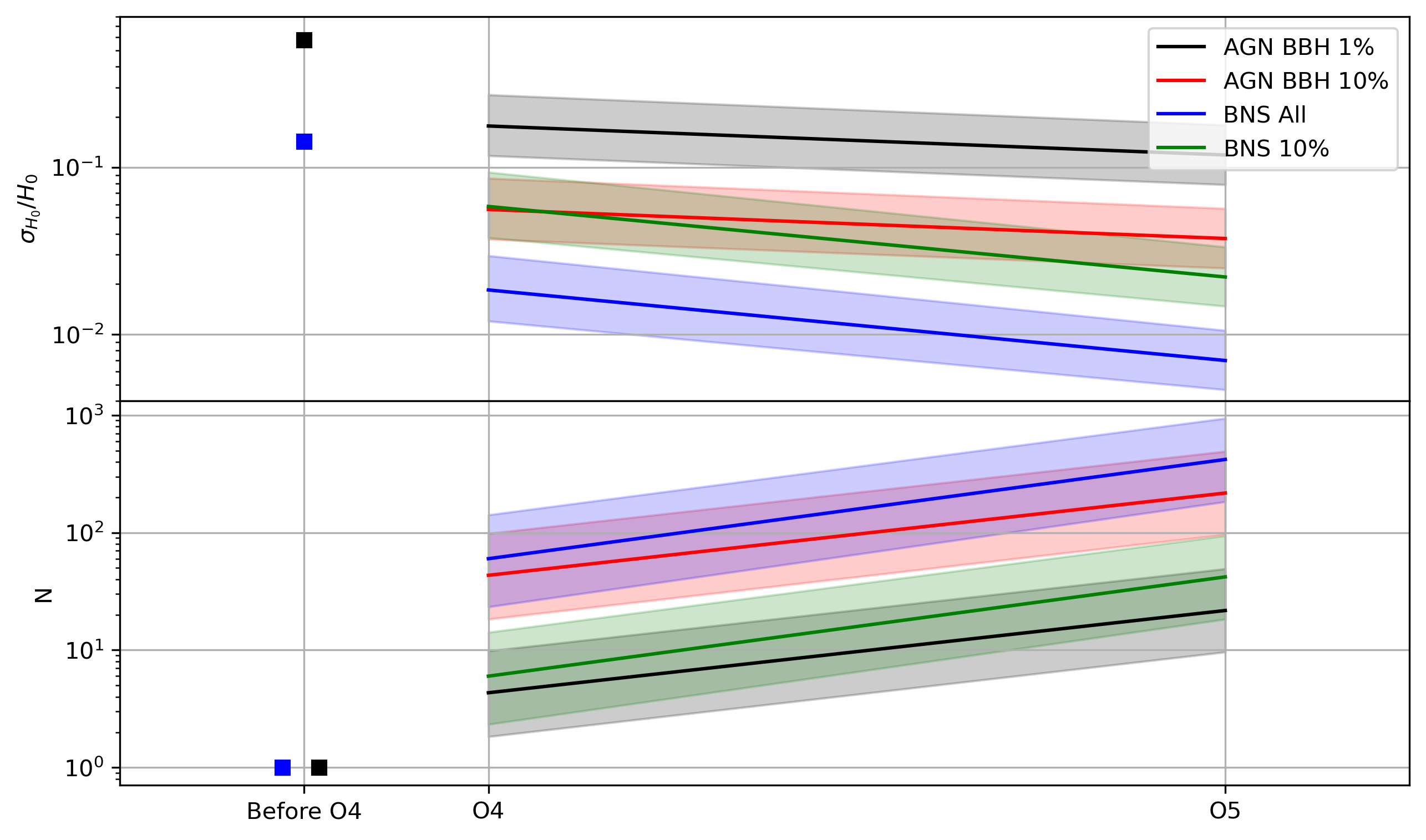} 
\caption{{\bf Projected relative uncertainty of Hubble constant measurements.} {\bf Top:} Relative uncertainties assuming that 1\% (black line) or 10\% (red line) of black hole mergers discovered by LIGO/Virgo are AGN-assisted and \textcolor{red}{have their redshift measured}. For comparison, we show relative uncertainties for neutron star mergers assuming that all (blue line) or only 10\% (green line) of them discovered by LIGO have identified host galaxies. The black square refers to the $H_0$ measurement from GW190521 \citep{GW190521_eBBH,gayathri2020hubble,2020arXiv200914199M,2022MNRAS.513.2152C} and the blue square, to the measurement from neutron star merger GW170817 \citep{Abbott_MMA_GW170817,2019NatAs...3..940H}. {\bf Bottom:} Expected number of detections for the same categories as above. \textcolor{red}{Following the color coding of the curves, the blue and black squares represent the one event employed in each existing measurement.}}
\label{fig:HubbleConst}
\end{figure}

Based on the estimate above, we considered two observation scenarios, in which 1\% and 10\% of LIGO/Virgo's black hole detections are both AGN-assisted and \textcolor{red}{have their redshift measured—through an electromagnetic counterpart or host galaxy identification}. For both scenarios, we computed the anticipated detection rate based on expected LIGO sensitivities in accordance \textcolor{red}{with \cite{emfollow}}. Our results are shown in Fig. \ref{fig:HubbleConst} (bottom). We see that by the end of this period, we expect about $N_{\rm gal}=20$ and $N_{\rm gal}=200$ detections for the 1\% and 10\% scenarios, respectively.

\textcolor{red}{Either association method—electromagnetic counterpart or host galaxy identification—has inherent uncertainty. We do not estimate this uncertainty, but rather assume that an application of our method to real detections would use electromagnetic/AGN association uncertainties as claimed in the literature. For a recent study of gravitational-wave measurements of the Hubble constant with uncertain galaxy identification, see \cite{alfradique2023dark}. Further discussion of the association between AGN and binary is beyond the scope of this paper.}

\textcolor{red}{It is worth noting that, in general, the inference of luminosity distance and other parameters from a binary black hole in an AGN can be impacted by systematic effects caused by the environmental material in the disk. AGN disk material impacts inspiralling binaries when they are in the LISA band. By the time they reach the LIGO band, however, the binary evolution decouples from the disk \citep{Toubiana_2021}, so environmental effects should not impact our proposed method for LIGO sources.}

\section{Measuring the Hubble Constant}

Using the expected $N_{\rm gal}$ shown in Fig. \ref{fig:HubbleConst} (bottom), we computed the expected uncertainty $\sigma_{H_0}$ with which we will be able to measure the Hubble constant. Our results are shown in Fig. \ref{fig:HubbleConst} (top). We see that within five years, with observing run O5, we expect to reach an uncertainty of $\sigma_{H_0}/H_0\approx$ \textcolor{red}{12}\% and \textcolor{red}{4}\% for our 1\% and 10\% models, respectively. This precision may be sufficient to help resolve (or deepen) the discrepancy between $H_0$ measurements using type Ia supernovae and the cosmic microwave background.

\textcolor{red}{For comparison, we show $\sigma_{H_0}/H_0$ expected for neutron star mergers. We assume $\sigma_{H_0}/H_0=1/7$ for a single BNS event, the uncertainty estimated from GW170817 \citep{GW170817_H0}. Because of the serendipity of GW170817, this is an optimistic choice for the single event BNS $H_0$ uncertainty.} For currently published detections by LIGO/Virgo, \textcolor{red}{50\% of neutron star mergers have redshift information from an electromagnetic counterpart}, but it is likely to drop further as more distant events are found by more sensitive gravitational-wave detectors. We therefore consider a realistic electromagnetic detection fraction of 10\%, as well as the optimistic case of 100\%. 

Our results for neutron star mergers are shown in Fig. \ref{fig:HubbleConst}. We see that if all neutron star mergers have a detected electromagnetic counterpart, \textcolor{red}{this constitutes the most precise case for the obtained $\sigma_{H_0}/H_0$}. For comparison, a similar calculation was carried out for neutron star mergers by \cite{2018Natur.562..545C} whose results are similar to ours assuming 100\% electromagnetic detection fraction. \textcolor{red}{If 10\% of both neutron star and AGN-driven black hole mergers have their redshift determined, then we find that these sources offer comparable precision for measuring $H_0$}. 

Compared to measurements of $H_0$ with type Ia supernova \citep{2019ApJ...876...85R} and the cosmic microwave background \citep{Planck:2018vyg}, those relying on gravitational waves (associated with electromagnetic counterparts) have had order 10 times more uncertainty \citep{GW170817_H0,gayathri2020hubble,2020arXiv200914199M,2022MNRAS.513.2152C}\textcolor{red}{, ranging from 20\% to 50\%}. The \textcolor{red}{channel} we propose may improve the accuracy of gravitational-wave-based measurements of the Hubble constant. With an increasing number of detections and well-localized events, starting with LIGO-Virgo-KAGRA's ongoing (as of this paper's writing) O4 run and improving in O5 and next-generation gravitational-wave-detector runs, the uncertainty provided by measurements of the Hubble constant with AGN-assisted black hole mergers should decrease more and more.

\textcolor{red}{In this paper, we have only considered the improvement in measuring the Hubble constant if BBHs can be localized to a single AGN, which can in turn provide the event's redshift. While a few events may be associated to a single AGN, localizing a BBH event to a handful of AGNs still provides redshift information, although with higher uncertainty, and allows for cosmological measurements. Considered together, a large number of BBHs localized to a handful of AGNs may provide an $H_0$ measurement with reasonable uncertainty and enhance the results reported here. Moreover, when an event involves objects in either mass gap, which are probable byproducts of hierarchical mergers and therefore likely inhabit AGNs, the mass information may be weighed in to strengthen the AGN association.} 

More research is needed to better understand possible multi-messenger emission mechanisms in BBH mergers in AGN disks.  Equally importantly, future BBH merger discoveries by LIGO-Virgo-KAGRA will need to be followed up by electromagnetic observatories, an effort that will need significant prioritization given the sheer rate of such detections. 

\section*{Acknowledgements}

The authors thank Karan Jani, Jon Gair, and Hsin-Yu Chen for useful suggestions. LMBA is grateful for the support of the Columbia Undergraduate Scholars Program and the Columbia Center for Career Education. AS is grateful for the support of the Stanford University Physics Department Fellowship. IB acknowledges support from the Alfred P. Sloan Foundation and the University of Florida. G.V. acknowledges the support of the
403 National Science Foundation under grant PHY-2207728. The authors thank Columbia University in the City of New York for its generous support. This research has made use of data, software, and/or web tools obtained from the Gravitational Wave Open Science Center (https://www.gw-openscience.org), a service of LIGO Laboratory, the LIGO Scientific Collaboration, and the Virgo Collaboration. 

\section*{Data Availability}

The data underlying this study will be made available upon reasonable request to the author.



\bibliographystyle{mnras}
\bibliography{BBH_AGN} 





\bsp	
\label{lastpage}
\end{document}